\documentclass[aip,apl,reprint]{revtex4-1}
\usepackage{graphicx}
\usepackage{dcolumn}
\usepackage{bm}
\usepackage{color,graphicx}
\usepackage{booktabs}

\begin{document}

\title{Tuning hole mobility in InP nanowires}

\author{M. Rebello Sousa Dias}
\affiliation{Departamento de F\'{i}sica, Universidade Federal de
S\~{a}o Carlos, 13565-905, S\~{a}o Carlos, S\~{a}o Paulo, Brazil}
\affiliation{Department of Physics and Astronomy and Nanoscale and
Quantum Phenomena Institute, Ohio University, Athens, Ohio
45701-2979}
\author {A. Picinin}
\affiliation{Departamento de F\'{i}sica, Universidade Federal de S\~{a}o Carlos, 13565-905, S\~{a}o Carlos, S\~{a}o Paulo, Brazil}
\author{V. Lopez-Richard}
\affiliation{Departamento de F\'{i}sica, Universidade Federal de S\~{a}o Carlos, 13565-905, S\~{a}o Carlos, S\~{a}o Paulo, Brazil}
\author{S. E. Ulloa}
\affiliation{Department of Physics and Astronomy and Nanoscale and Quantum Phenomena Institute, Ohio University, Athens, Ohio 45701-2979}
\author{L. K. Castelano}
\affiliation{Departamento de F\'{i}sica, Universidade Federal de S\~{a}o Carlos, 13565-905, S\~{a}o Carlos, S\~{a}o Paulo, Brazil}
\author {J. P. Rino}
\affiliation{Departamento de F\'{i}sica, Universidade Federal de S\~{a}o Carlos, 13565-905, S\~{a}o Carlos, S\~{a}o Paulo, Brazil}
\author{G. E. Marques}
\affiliation{Departamento de F\'{i}sica, Universidade Federal de S\~{a}o Carlos, 13565-905, S\~{a}o Carlos, S\~{a}o Paulo, Brazil}

\begin{abstract}

Transport properties of holes in InP nanowires were calculated
considering electron-phonon interaction via deformation potentials,
the effect of temperature and strain fields. Using molecular
dynamics, we simulate nanowire structures, LO-phonon energy
renormalization  and lifetime. The  valence band ground state
changes between light- and heavy-hole character, as the strain
fields and the nanowire size are changed.  Drastic changes in the
mobility arise with the onset of resonance between the LO-phonons
and the separation between valence subbands.
\end{abstract}

\maketitle

Semiconductor nanowires (NWs) are increasingly used in a wide range of devices. They appear as building blocks of
nanocircuits\cite{nanocircuit} and can be applied in electrically driven lasing\cite{naturelaser}, which can be used in telecommunications
and information storage for medical diagnostics and therapeutics\cite{lasrapplication}. Improvements in NW synthesis, including chemical
technics, allow thorough control of their shape, size and composition \cite{science279, lieber1, lieber2, bjork, nano} along with detailed
microscopic characterization of built-in strain fields.\cite{nano, prb}  As the conductivity is mostly defined by the carrier-phonon
interaction and phonon-lifetime, tuning of the NW structural properties could result in the possibility of also finding optimal conditions
for carrier transport.

Considerable efforts have been devoted to the description of carriers in the conduction band of
NWs,\cite{cond0,cond1,cond2,cond4,cond5,cond6} while similar endeavors are not so common for holes in the valence band.\cite{val1} As the
mobility is inversely proportional to the carrier effective mass, one may naturally expect that considering carriers in the valence band
may result in a drop in mobility when compared to the light electrons in the conduction band. This could certainly be the case for
heavy-hole (hh) transport; however, light-holes (lh) under certain conditions may be promoted to be the top valence band by tuning
structural parameters of NWs.\cite{mariama} This atypical circumstance is the result of confinement effects and hh-lh mixing, affected as
well by strain and surface asymmetry fields.\cite{nano, mariama} As we will show here, this results in significant mobility enhancement
for lh in suitable NWs.  We can also take advantage of valence band mass anisotropy to attain resonant conditions that allow sharp
variations of the hole mobility with external parameters, especially when the leading scattering process involve longitudinal optical
phonons (LO-phonons) through the deformation potential.\cite{Raman2} Additional hole-phonon interactions,\cite{mahan} such as deformation
potential and piezoelectric coupling to acoustic phonons and polar coupling to optical phonons,\cite{Raman1} have weaker effects and will
not be consider here.\cite{mahan,cardonabook} In order to provide realistic estimates of the expected mobility changes in the NWs of
interest, we consider the effects of dimensionality reduction on the LO-phonon dispersion and lifetime, using molecular dynamics
simulations for different NWs size and at various temperatures.

We consider different NW cross sections and shapes, while temperature effects are included in the mobility through the phonon occupation
and strain effects in a multiband Luttinger Hamiltonian.  We find that mobility changes in a non-monotonic fashion according to NW width,
strain fields, and temperature. In particular, we show that for certain NW widths, one finds resonant behavior that greatly suppresses the
hole mobility and is strongly affected by temperature and strain. The interaction potentials used in our molecular dynamics (MD)
simulations consist of two- and three-body interaction terms, as described by Branicio \emph{et
al.}\cite{ref-md1,ref-md2,ref-md3,ref-md4,ref-md5}. The parameters of the interatomic potential are determined using the cohesive energy,
density, bulk modulus and elastic constant $C_{11}$ of the material as described before,\cite{ref-md5} with some slight
adjustments.~\cite{foot}

\begin{figure}[hbt]
\linespread{1.0}
\begin{center}
\includegraphics[scale=0.95]{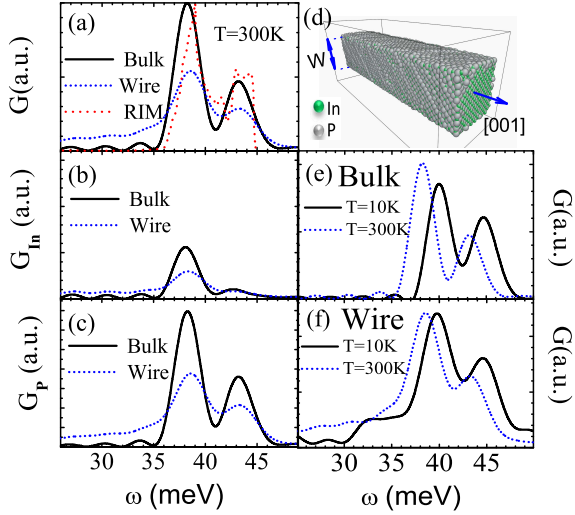}
\end{center}
\vspace{-0.65cm}\caption{ VDOS for NW and bulk. (a) Total VDOS; rigid ion model (RIM) VDOS for bulk from data.\cite{ref-md7} (b) and (c)
Partial VDOS for bulk and NW at $300 K$: (b) indium contribution and (c) phosphorus contribution. (e) and (f) VDOS for bulk and NW at 10 K
and 300 K. (d) Simulated InP NW structure by molecular dynamics at $T=300 K$. Green (grey) dots represent Indium (Phosphorous) atoms.}
\label{wire}
\end{figure}

This potential provides excellent estimates for melting temperature, structural phase transformation induced by pressure, and specific
heat\cite{ref-md2} and describes well the vibrational density of states of the material. We obtain the phonon density of states in InP NWs
considering different temperatures. The NWs were created cutting a block of a perfect crystal with the z-axis along the [001] direction
with periodic boundary conditions in the z-direction. The x- and y-directions were surrounded by a vacuum region. The system consisted
typically of nine unit cells along x- and y-directions and forty unit cells along the z-direction ($53 \mathring{A} \times 53 \mathring{A}
\times 234.5 \mathring{A}$); the total number of atoms is 25,920 (12,960 In + 12,960 P) (Fig.~\ref{wire}(d)).  The NW is allowed to relax
during a long simulation run (25,000 time steps, one time step=1.5fs) at each temperature. After this relaxation time, a few surface
defects can be observed.

The velocity-velocity auto correlation function, $Z\left( t\right) =\frac{\left\langle \vec{v}_{i}\left( t\right) .%
\vec{v}_{i}\left( 0\right) \right\rangle }{\left\langle \vec{v}_{i}\left( 0\right) .\vec{v}_{i}\left( 0\right) \right\rangle }$, where
$\vec{v}_{i}\left( t\right)$ is the velocity of particle $i$ at time $t$ and the brackets are averages over ensembles and particles. The
vibrational phonon density of states (VDOS) is determined through the Fourier transform \cite{ref-md8}
\begin{equation}
G_{l}\left( \omega \right) =\frac{6N_{l}}{\pi }\int_{0}^{\infty }Z_{l}\left( t\right) \cos \left( \omega t\right) dt,
\end{equation}
where the subindex $l$ is the atom In or P.

Fig.~\ref{wire}(a) compares the VDOS computed from MD calculations for a bulk system (solid curve) to the VDOS extracted from the rigid
ion model (dashed curve) based on experimental results\cite{ref-md6,ref-md7}. The MD results reproduce very well the main characteristics
of the experimental, predicting the existence of the transversal optical (TO, $\sim43meV$) and longitudinal optical (LO, $\sim38 meV$)
modes, and a gap between $22 meV$ and $36 meV$ (acoustic modes below $22 meV$ not shown).

The effect of the surfaces on the VDOS of the NW is also shown in Fig.~\ref{wire}(a)-(c). The main NW characteristics in the VDOS resemble
the bulk results; however, some differences can be observed. The NW surfaces appreciably increases the amount of modes in the gap region,
between $22 meV$ and $36 meV$. Although the TO and LO modes are dominant, the NW surface plays an important role by inducing surface modes
that appear in the gap region. We notice that the P atoms make the main contribution to the optical modes in the gap have a larger VDOS,
and an overall large mode width for the NW. The optical modes are characterized by the relative displacement between ions and the lighter
atoms usually dominate such modes, as shown. Moreover, we observe a slight blue shift of the LO mode ($\sim 1meV$) with respect to the
bulk.

The effect of temperature on the VDOS for both NW and bulk structure is shown in Figs.~\ref{wire}(e) and (f). We see a general broadening
of the modes and a shift in the peak position to lower frequencies with increasing temperature. Although the gaps region keep a similar
profile, the VDOS in the gaps increase with temperature. We notice also that the temperature affects the contrast between LO and TO modes,
especially in the NW, Fig.~\ref{wire}(f), by increasing the LO amplitude.

Notice that the mode broadening is slightly weaker for the NW than for the bulk. Based on the MD results, we verify that the TO and LO
modes are the most important vibrational modes. Also, the NW exibits a shift in the LO mode with temperature, which is an important fact
to be considered when calculating the contribution of hole-phonon scattering to the mobility of NWs.

The valence band Hamiltonian for the NW can describe confinement effects, mass anisotropy, and strain fields within the same framework, as
one writes,\cite{luttinger}
\begin{equation}
\mathcal{H}_{hh}=-\left(\frac{\gamma_{1}+\gamma_{2}}{2}\right)\{\hat{k}_{+},\hat{k}_{-}\}-\left(\frac{\gamma_{1}-2\gamma_{2}}{2}\right)\hat{k}_{z}^{2},
\end{equation}
\begin{equation}
\mathcal{H}_{lh}=-\left(\frac{\gamma_{1}-\gamma_{2}}{2}\right)\{\hat{k}_{+},\hat{k}_{-}\}-\left(\frac{\gamma_{1}+2\gamma_{2}}{2}\right)\hat{k}_{z}^{2},
\end{equation}
for the heavy- and light-holes, where $\gamma_{\alpha}$ ($\alpha=1,2,3$) are the Luttinger parameters, $\{A,B\}=\frac{1}{2}(AB+BA)$, and
$\hat{k}_{\pm}=\hat{k}_{x}\pm i\hat{k}_{y}$. Notice that the subband with hh character along the wire has a {\em low} effective mass in
the transverse direction $\approx(\gamma_{1} + \gamma_{2})^{-1}$, while the lh subband has a {\em large} transverse mass
$\approx(\gamma_{1} - \gamma_{2})^{-1}$; the different transverse masses result in the possible inversion of the lh and hh subband
ordering, due to the NW confinement effects. Strain effects lead to modulation of the valence subbands,\cite{nano} introducing a subband
displacement given by \cite{cardona} $\Delta \mathcal{H}_{hh}=-P+Q$ and $\Delta \mathcal{H}_{lh}=-P+Q+\frac{2Q^{2}}{\Delta_{so}}$, where
$P=2(a_{v}+a_{c})(\frac{c_{11}+c_{12}}{c_{11}})\varepsilon_{||}$, $Q=-b(\frac{c_{11}+2c_{12}}{c_{11}})\varepsilon_{||}$ and
$\Delta_{so}=0.108 eV$ is the spin-orbit split-off energy.~\cite{foot1} The hole wave function in the NW has the form
$|\Psi_{i}\rangle=|\psi_{i}\rangle|J,m_{j}\rangle$, where $|\psi_{i}\rangle$ is the envelope function, which depends on the cross section
of the NW, and $|J,m_{j}\rangle$ is the total angular momentum eigenstate, $|3/2,\pm 3/2\rangle$ for pure hh character, and
$|3/2,\pm1/2\rangle$ for the lh.

The hole-phonon interaction Hamiltonian is given by
\begin{eqnarray}
\mathcal{H}_{h-p}=\sum_{\mathbf{q}}M_{\mathbf{q}}U_{h-i}(\mathbf{q})[\hat{a}_{\mathbf{q}}e^{i \mathbf{r}.\mathbf{q}}+\hat{a}_{\mathbf{q}}^{\dag}e^{-i \mathbf{r}.\mathbf{q}}],
\label{interac}
\end{eqnarray}
with $M_{\mathbf{q}}=(\mathbf{q}\cdot\mathbf{\varepsilon}_{q})(\hbar/2\rho\omega_{\mathbf{q}}V)^{\frac{1}{2}}$, where $\mathbf{q}$ is the
phonon wave vector for polarization vector $\mathbf{\varepsilon}_{q}$, $\rho$ is the mass density, $V$ is the system volume, and
$U_{h-i}(\mathbf{q})$  is given in terms of the deformation potential for holes.\cite{annacross, cardonabook} Considering long wavelength
processes, we have $U_{h-i}(\mathbf{q})\propto\mathbf{u}$, where $\mathbf{u}$ is the relative displacement between atoms inside the
primitive unit cell. By symmetry, states with hh character couple with those of lh character along the direction $[001]$\cite{cardonabook,
annacross}. As this coincides with the wire axis, along which the carrier transport takes place, we find $ \langle
hh^{\pm}|U_{h-i}(z)|lh^{\mp}\rangle=\langle lh^{\mp}|U_{h-i}(z)|hh^{\pm}\rangle=\frac{\pm i d_{0}}{2 a_{0}}$, where $d_{0}$ is the
deformation constant and $a_{0}$ the lattice parameter.~\cite{foot2}

The mobility is given by $\mu=\frac{e}{m_{0}\lambda_{z\beta}}\tau$, in terms of the hole-phonon scattering time, $\tau^{-1}
=\sum_{q}S(k,q)$, and the transition rate
\begin{eqnarray}\label{FGR}
S(k,k')=\frac{2 \pi}{\hbar}\left[|\langle \mathcal{H}_{h-p}^{a}\rangle|^{2}\delta(E_{f}(k')-E_{i}(k)-\hbar\omega_{\mathbf{q}})+ \right. \nonumber \\
\left. |\langle\mathcal{H}_{h-p}^{e}\rangle|^{2}\delta(E_{f}(k')-E_{i}(k)+\hbar\omega_{\mathbf{q}})\right],
\end{eqnarray}
where $k$ and $k'$ refers to the initial and final states, and $\mathcal{H}_{h-p}^{a}$ and $\mathcal{H}_{h-p}^{e}$ refer to the phonon
absorption and emission processes in Eq.~(\ref{interac}). The phonon density is assumed to be given by a Lorentzian centered at
$\omega_{LO}$ with width $\gamma$.  Both of these values shift with temperature, as discussed in the previous section. We may now analyze
the effects of strain and temperature on the hole mobility.

\begin{figure}[hbt]
\linespread{1.0}
\begin{center}
\includegraphics[scale=1]{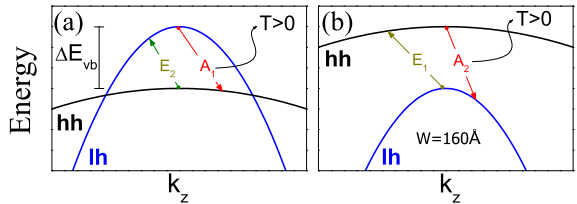}
\end{center}
\vspace{-0.75cm}\caption{Valence band ground states for NW of width W. (a) When the lh occupies the ground state. (b) When the hh occupies
the ground state.} \label{valenceband}
\end{figure}
To characterize the initial  and final states involved in the scattering processes that affect the mobility, we show in
Fig.~\ref{valenceband} the relevant valence band structure for two different cases. For thin NWs, with or without strain, the finite NW
width leads to a picture similar to Fig.~\ref{valenceband}(a), where the lh subband is promoted to the top given its higher transverse
effective mass, as discussed before. Thus, under such conditions, a hh can be scattered to the subband with lh character through phonon
emission (process E$_{1}$), and at $T>0$ the lh can be excited to the hh subband via phonon absorption (process A$_{1}$). In the presence
of lateral compressive strain, the subbands may switch their relative positions with the hh assuming the top at large NW width. Then, a lh
might be scattered via phonon emission (process E$_{2}$) while a hh can be affected by phonon absorption at $T>0$ (process A$_{2}$).
Notice that by changing the wire radius, one can reach a resonant condition ($\Delta E_{vb}=\hbar\omega_{LO}$). On the other hand, with
strain, depending on the value of the NW width, the ground state can have a character lh (thin NW) or hh (thick). This behavior is similar
for all NW cross sections, as it reflect the transversal quantization.
\begin{figure}[hbt]
\linespread{1.0}
\begin{center}
\includegraphics[scale=0.95]{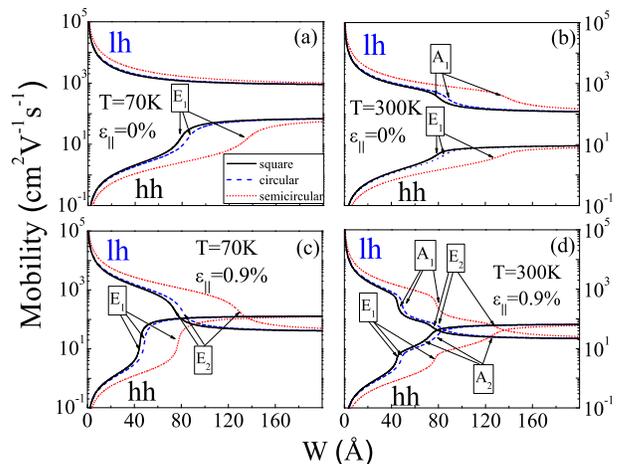}
\end{center}
\vspace{-0.75cm}\caption{Hole mobility versus NW width W for states with $k_{z}=0$. (a) System at T$=70$K without strain. (b) System at
T$=300$K without strain. (c) System at T$=70$K  with strain. (d) System at T$=300$K  with strain.} \label{mob1}
\end{figure}

The relative position of the valence subbands is extremely important for the carrier transport in NWs. Given the mobility dependence on
the longitudinal effective mass, which modulates the hole-phonon interaction, valence subband shifts may produce sharp fluctuations of the
mobility as temperature or structural parameters change. Fig.~\ref{mob1} shows the mobility for different strain and temperature values as
function of the NW width. In Figs.~\ref{mob1}(a)-(b), the mobility reflects a band configuration similar to the one depicted in
Fig.~\ref{valenceband}(a). At T$=70K$, in Fig.~\ref{mob1}(a), the increase effect of phonon absortion leads to the monotonic decrease of
the lh mobility whit increasing NW width, while the hh displays a monotonic mobility increase, as the intersubband separation decreases
with increasing NW  width. Also, a sharp variation near the region where $\Delta E_{vb}\sim \hbar \omega_{LO}$ is seen, the resonant
condition greatly enhances LO phonon emission by a hh in panel (a). At higher temperatures, Fig.~\ref{mob1}(b), the resonant condition
also affects the carriers in the lh subband, producing a sharp drop in mobility (from A$_{1}$ processes).

Given the band structure modulation with strain, the condition $|\Delta E_{vb}|\sim \hbar \omega_{LO}$ can be attained twice by varying
the NW width (corresponding to the cases displayed in Figs.~\ref{valenceband}(a) and (b)). Thus, two resonant regions appear in
Fig.~\ref{mob1}(c) where phonons can be emitted by both the hh and lh subbands (E$_{1}$ an E$_{2}$ processes, respectively). At higher
temperatures, the phonon absorption features appear as additional jumps in the mobility, shown in Fig.~\ref{mob1}(d), processes A$_{1}$
and A$_{2}$. Notice that the lh and hh subband inversion with increasing NW width, in the presence of strain, is accompanied by crossing
of the mobility curves, Figs.~\ref{mob1}(c)-(d). Tuning the mobility of a hole system via in-situ changes of the NW width or strain
fields, is not an easy task in experiments. As we will see below, however, one can achieve drastic in-situ mobility changes for NWs close
to the resonance condition by suitable changes in temperature.
\begin{figure}[hbt]
\linespread{1.0}
\begin{center}
\includegraphics[scale=1]{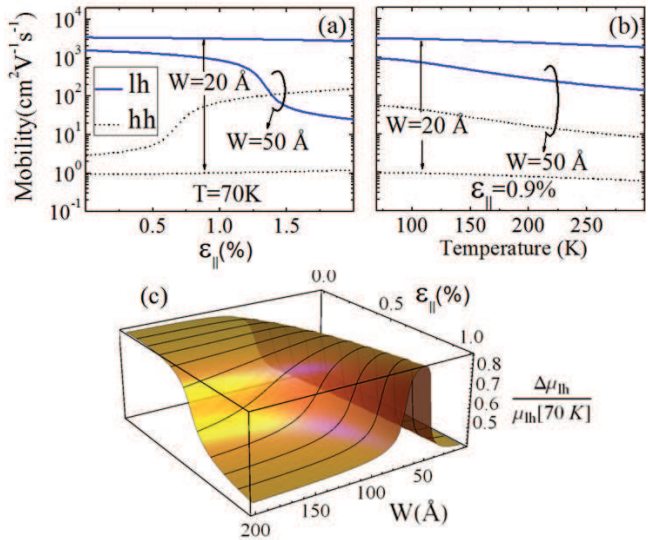}
\end{center}
\vspace{-1.3cm}\caption{(a) and (b) hole mobility for two values of NW width for states with $k_{z}=0$: (a) as function of strain for
T$=70$K, and (b) as function of temperature for $\varepsilon_{||}=0.9\%$. (c) Light hole mobility ratio, where
$\Delta\mu_{lh}=\left(\mu_{lh}[70K]-\mu_{lh}[300K]\right)$, versus wire width for different values of strain at $k_{z}=0$.} \label{mob3}
\end{figure}

In Fig.~\ref{mob3}(a) for a square NW width of $W=20\mathring{A}$, no resonant signatures appear in the strain and temperature range
analyzed, while for thicker NWs such conditions become evident for both kind of holes. This situation follows the trends described in
Fig.~\ref{valenceband}, since for thinner NWs, the valence band configuration corresponds to that in Fig.~\ref{valenceband}(a), with
$\Delta E_{vb}>\hbar\omega_{LO}$. In turn, by raising the temperature, the process of phonon absorption becomes more effective, reducing
the lh mobility at higher temperatures; as displayed in Figs.~\ref{mob3}(b). The effect is present for all NWs, although with different
features for various $W$ values, as we now discuss.

Fig.~\ref{mob3}(c) shows the strongly non-monotonic mobility variation with temperature for certain NW widths with energies close to the
resonant condition, $\Delta E_{vb} \simeq \hbar \omega_{LO}$. This figure shows the lh mobility ratio of high and low temperatures,
$\Delta\mu_{lh}/\mu_{lh}[70K]$, where $\Delta\mu_{lh}=\left(\mu_{lh}[70K]-\mu_{lh}[300K]\right)$, as function of the NW width for
different values of strain. For lh-holes, the mobility at low temperatures is high (Fig.~\ref{mob3}(b)) and according to the radius and
strain one may observe a drastic drop in mobility at high temperature. For a system with no strain, for example, a large but monotonic
drop in the mobility is seen for NWs with large width (larger than $100\mathring{A}$). In real systems, however, free standing NWs grow
with built-in strain,\cite{nano} which has a direct impact on the dependence of $\mu$ on the temperature and width. As the strain
increases, the lh-hh subband reversal is possible as the NW width increases, resulting in highly sensitive mobility on temperature and/or
width. For example, for $\varepsilon_{||}=0.9\%$,~\cite{nano} and for width values close to $50\mathring{A}$ mobility exhibit a sharp
change in the mobility with temperature.

The authors acknowledge the support of CAPES, CNPQ, FAPESP, NSF, and MWM/CIAM.


\begin{thebibliography}{99}
\bibitem{foot1}For deformation potentials, in (dyn cm$^{-2}$) units, we have $c_{11}=10.11\times10^{11}$ and $c_{12}=5.61\times10^{11}$. For elastic modules in (eV) $a_{c}=-7.0$, $a_{v}=-0.6$, and $b=-1.7$.~\cite{Landolt,PRL} When used, the mean value of the biaxial strain was $\varepsilon_{||}=0.9\%$.~\cite{nano}
\end{thebibliography}

\end{document}